\documentclass[10pt]{article}
\usepackage{amsmath}
\usepackage{amsfonts}

\usepackage{amsmath,amssymb}
\usepackage{graphicx}
\usepackage{color}
\usepackage{url}

\newtheorem{theorem}{Theorem}
\newtheorem{lemma}{Lemma}
\newtheorem{corollary}{Corollary}

\newtheorem{proposition}{Proposition}
\newtheorem{remark}{Remark}
\newtheorem{conjecture}{Conjecture}

\newtheorem{example}{Example}

\newcommand{\RNum}[1]{\uppercase\expandafter{\romannumeral #1\relax}}

\newcommand{\F}{\ensuremath{\mathbb F}}

\newcommand{\done}{\hfill $\Box$ }

\newcommand{\Tr}{{{\rm Tr}}}

\newcommand{\notequiv}{{\,\not\equiv\, }}

\newcommand{\ls}[1]
    {\dimen0=\fontdimen6\the\font\lineskip=#1\dimen0
     \advance\lineskip.5\fontdimen5\the\font
     \advance\lineskip-\dimen0
     \lineskiplimit=0.9\lineskip
     \baselineskip=\lineskip
     \advance\baselineskip\dimen0
     \normallineskip\lineskip\normallineskiplimit\lineskiplimit
     \normalbaselineskip\baselineskip
     \ignorespaces}

\begin{document}

\bibliographystyle{abbrv}

\title{More Classes of  Complete Permutation Polynomials over $\F_q$}

\author{Gaofei Wu
\thanks{G. Wu is with the State Key Laboratory of Integrated Service Networks,
Xidian University, Xi'an, 710071, China.
He is a visiting PhD student (Sep. 2012- Aug. 2014) in the Department of Informatics, University of Bergen.
 Email: gaofei\_wu@qq.com.},
Nian Li
\thanks{N. Li is with the Information Security and National Computing Grid Laboratory,
Southwest Jiaotong University, Chengdu, 610031,
 China.
 Email:
nianli.2010@gmail.com.},
Tor Helleseth
\thanks{T. Helleseth is with the  Department of Informatics, University of Bergen,
 N-5020 Bergen, Norway. Email:
Tor.Helleseth@ii.uib.no.},
and
Yuqing Zhang
\thanks{Y. Zhang  is with the  National Computer Network Intrusion Protection Center, UCAS, Beijing 100043, China.
 Email:
zhangyq@ucas.ac.cn.}}
\date{}
\maketitle

\thispagestyle{plain} \setcounter{page}{1}

\begin{abstract}
In this paper,
by using a powerful criterion for permutation polynomials  given by Zieve, we give several classes of
complete permutation  monomials  over $\F_{q^r}$.
In addition, we present a class of complete permutation multinomials, which is a generalization of
 recent work.
%

%


\end{abstract}

\ls{1.5}
\section{Introduction}

Let $p$ be a prime and $n$ be a positive integer.
Let $\F_q$  be a finite field of $q=p^n$ elements. We denote
$\F_q^* $ the multiplication group of $\F_{q}$.
A polynomial $f\in \F_{q}[x]$ is called a permutation polynomial (PP) if
the associated polynomial mapping
$f:c\mapsto f(c)$ from $\F_{q}$ to itself is a permutation over
$\F_{q}$ \cite{Lidl97}.
Permutation polynomials over finite fields  have important applications
in cryptography, coding theory, and combinatorial design theory.
There has been lots of results about  PPs over
$\F_{q}$ \cite{Akbary11,Charpin03,Coulter09,Ding09,Dobbertin98,Li13,Lidl88,Lidl93,Marcos11,Shallue13,YuanDing08,Zieve09}.

A polynomial $f\in \F_{q}[x]$ is called a complete  permutation polynomial (CPP) if
 both $f(x)$ and $f(x)+x$ are permutations over $\F_{q}.$
For some known results  on  CPPs over $\F_{q}$  see \cite{Laigle07,Mullen87,Nied82,Wan86,Yuan07}.
For a positive integer $d$ and $a\in \F_{q}^*,$ a monomial function
$ax^d$ is a CPP over $\F_{q}$ if and only if $\gcd(d,q-1)=1$ and $ax^d+x$ is a PP
over $\F_{q}$,
such a   $d$ is called  a CPP exponent over $\F_{q}.$

Recently,  some classes of  CPPs over $\F_{p^n}$ are given in \cite{Tu13,Tu132,Bao13, Wu13,Xu13,Zieve13,Zieve132}.
In this paper,   we give some classes of
CPPs over $\F_{p^n}$ as follows:
\begin{enumerate}

\item[(1)] For $p=3,$ $n=2k, $ and $ d=p^k+2,$
we prove that  $d$ is a CPP exponent  over $\F_{p^n}$.

\item[(2)] For any odd prime $p$,  $n=2k, $ and $ d=(p^k-1)\cdot \frac{p^i-1}{2}+p^i$   ($1\leq i\leq n$),
we prove that  $d$ is a CPP exponent  over $\F_{p^n}.$

\item[(3)] For any odd prime $p$, $d=\frac{p^{rk}-1}{p^k-1}+1,$
for $r=4$, we give a sufficient and necessary condition for $a^{-1}x^d$
to be a CPP over $\F_{p^n},$ where $a\in \F^*_{p^k}.$ For
$r=6$ and $p=3 $ or $p=5$, we show that
$d$ is a CPP exponent over $\F_{p^{rk}}$, where $\gcd(r,k)=1.$

\item[(4)]For any prime $p$,   $n=rk,$ and  $f(x)=x(\Tr_{k}^{n}(x))^{p-1}+(p-1)x^{p}+ax,\, a\in\F_{p^k}\setminus\{0,-1\}$,
we prove that  $f(x)$ is a CPP over $\F_{p^n} $ if
$\gcd(p,r)=\gcd(p-1,r)=1,$
 where $\Tr_{k}^n(x)$ is the trace function from $\F_{p^{n}}$ to
$\F_{p^k}.$
\end{enumerate}


The first two classes  of monomial CPPs are with  Niho exponents.
A positive integer $d$ (always understood modulo $p^n-1$) is a
Niho exponent if $d\equiv p^j \,({\rm mod}\, p^n-1)$ for
some $j<n$ \cite{Niho72}.
Note that the first class is a special case of \cite[Corollary 3.4]{Zieve132}, and also has been proved in \cite[Theorem 3.1]{Xu13}
for $k$ odd.
The  first class is  also  a special case of the second class with   $p=3$ and $i=1$.
Here we list it separately is because that  the
proofs of these two classes of CPPs are different, and
the proof of $3^k+2$ to be a CPP exponent
over $\F_{3^{2k}}$
 is very  interesting\footnote{The method we used to prove that $3^k+2$ is  a CPP exponent
  is different from the ones used
 in \cite{Xu13,Zieve132}.}.
The third class is a further study of \cite{Wu13} and the fourth class is a generalization of the main result
in \cite{Bao13}.

\section{Preliminaries}

Let $p$ be  prime, $n,\, k$ be two integers such that
$k|n$. The trace function from
$\F_{p^n} $ onto $\F_{p^k}$ is defined by
$$
\Tr_k^n(x)=\sum_{i=0}^{n/k-1}x^{p^{ik}}, \, x\in \F_{p^n}.
$$

The Walsh transform of  a function $f$ from
$\F_{p^n}$ to $\F_p$ is defined by
$$
W_{f}(a)=\sum_{x\in\F_{p^n}}\omega^{f(x)+\Tr_1^n(ax)},
$$
where $a\in \F_{p^n}$, and  $\omega$ is the complex primitive $p$-th root of unity.

In \cite{Lidl97}, a criterion for PPs is given by using the
additive characters of the underlying finite field.

\begin{lemma}\label{lemma}\cite{Lidl97}
A mapping  $g:\F_{p^n}\rightarrow \F_{p^n}$ is a PP if and only if
 for every $\alpha\in \F_{p^n}^*$,
 $$
 \sum_{x\in  \F_{p^{n}}}\omega^{\Tr_1^n(\alpha g(x))}=0.
 $$
\end{lemma}


The following lemmas will also be needed in the sequel.
\begin{lemma}\cite{Lidl97}\label{irre}
An irreducible polynomial over $\F_q$ of degree $n$ remains  irreducible over $\F_{q^k}$
if and only if $\gcd(k,n)=1$.
\end{lemma}


\begin{lemma}\cite{Zieve09}\label{lemkey}
Let $p$ be a prime.  Let $l,\, n $ and $s$ be positive integers such that $s|p^n-1.$
Let $g(x)\in \F_{p^n}[x].$
Then $f(x)=x^lg(x^{\frac{p^n-1}{s}})$ is a PP over $\F_{p^n}$ if and only if
$\gcd(l,\frac{p^n-1}{s})=1$ and
 $x^lg(x)^{\frac{p^n-1}{s}}$ is a permutation of $\mu_s,$
 where $\mu_s$ is the set of $s$-th roots of unity in $\F_{p^n}.$
\end{lemma}


\begin{lemma}\cite[Theorem 1.1]{Zieve132}\label{lemkeyzieve}
Let $p$ be a prime.  Let $l,\,r,\,k$ and $ n $  be positive integers such that $n=rk.$
Let $g(x)\in \F_{p^n}[x].$
Then $f(x)=x^lg(x^{\frac{p^n-1}{p^k-1}})$ is a PP over $\F_{p^n}$ if and only if
$\gcd(l,\frac{p^n-1}{p^k-1})=1$ and
 $$x^lg(x)g^{p^k}(x)g^{p^{2k}}(x)\cdots g^{p^{(r-1)k}}(x)$$
  is a permutation of $\F_{p^k}$, where
  $g^{p^{ik}}(x)$ denotes the polynomial obtained from $g(x)$ by raising every coefficient to the
  $p^{ik}$-th power.
\end{lemma}
In \cite{Zieve132}, Zieve gave some classes of CPPs over $\F_{p^{rk}}$ by using  Lemma \ref{lemkeyzieve} and some PPs of degree
$d\leq 5$ over $\F_{p^k}$.

Let $n,\,r,\, k$ be integers such that $n=rk$.
For any $a\in \F_{p^n},$ let  $a_i=a^{p^{ik}},$ where  $0 \leq i\leq r-1$.
Define
\begin{eqnarray}
h_a(x)=x\prod_{i=0}^{r-1}(x+a_i)= x\sum_{i=0}^{r}\lambda_i x^{r-i},\label{hay}
\end{eqnarray}
where $\lambda_0=1, $ and $\lambda_i=\sum_{0\leq j_1<j_2<\cdots<j_i\leq r-1}a_{j_1}a_{j_2}\cdots a_{j_i}$
for $ 1\leq i \leq r. $
It is easily seen  that $\lambda_i\in\F_{p^k}$ for $0\leq i \leq r.$ Thus
 $h_a(x)\in\F_{p^k}[x]$.

 Let $ l=1 $ and $g(x)=x+a\in \F_{p^n}[x]$ in Lemma \ref{lemkeyzieve}, we have

\begin{lemma}\footnote{see also in \cite[Lemma 6]{Wu13} for the case  $p=2$.}\label{p3nrk}
 Let $n=rk,$  and  $d=\frac{p^{rk}-1}{p^k-1}+1$.
Then  $x^d+ax\in\F_{p^n}[x]$ is a PP over $\F_{p^n}$
   if and  only if
   $h_a(x)\in\F_{p^k}[x]$ is a PP over $\F_{p^k}$.
\end{lemma}



The PPs of low degree  over $\F_{q}$ will be used in the  sequel.
A normalized degree $l$ polynomial $f(x)\in\F_{q}[x]$ is a monic polynomial
  that
  satisfies the following
 conditions:
 $f(0)=0$, and if $\gcd(l,q)=1$, the coefficient of $x^{l-1} $ is $0$.
 All normalized PPs  of degree
$\leq 5$ are listed
in \cite[Table 7.1]{Lidl97}. For the reader's convenience, we copy it here  in Table
\ref{ppdeg5}.


\begin{table}[h!]
  \begin{center}
  \caption{Normalized PPs $f(x)$ of degree $\leq 5$ over $\F_q$}
\label{ppdeg5}
    \begin{tabular}{| l| c |}
    \hline
     $f(x)$      &   $\F_q$   \\
    \hline\hline
   $x$      &  any $q$   \\
    \hline
       $x^2$      &  $q\equiv 0\, ({\rm mod }\, 2)$  \\
    \hline
       $x^3$      &  $q\notequiv 1\, ({\rm mod }\, 3)$  \\
    \hline
       $x^3-vx$ ($v$ not a square)      &  $q\equiv 0\, ({\rm mod }\, 3)$  \\
    \hline
       $x^4\pm 3x$      &  $q=7$   \\
    \hline
       $x^4+v_1x^2+v_2x$ (if $0$ is its only root in $\F_q$)      &  $q\equiv 0\, ({\rm mod }\, 2)$  \\
    \hline
       $x^5$      &  $q\notequiv 1\, ({\rm mod }\, 5)$  \\
    \hline
       $x^5-vx$  ($v$ not a fourth power)    &  $q\equiv 0\, ({\rm mod }\, 5)$  \\
    \hline
       $x^5+vx$  ($v^2=-1$)    &  $q=9$  \\
    \hline
         $x^5\pm 2x^2$    &  $q=7$  \\
    \hline
         $x^5+vx^3\pm x^2 +3v^2x$  ($v$ not a square)    &  $q=7$  \\
    \hline
         $x^5+vx^3 +5^{-1}v^2x$     &  $q\equiv \pm 2\, ({\rm mod }\, 5)$  \\
    \hline
         $x^5+vx^3+3v^2x$  ($v$ not a square)    &  $q=13$  \\
    \hline
         $x^5-2vx^3 +v^2x$  ($v$ not a square)    &  $q\equiv 0\, ({\rm mod }\, 5)$  \\
    \hline
         $x^5+x^2 +x$      &  $q=2$  \\
    \hline
         $x^5+x^3 +x^2$      &  $q=2$  \\
    \hline
         $x^5 +x$      &  $q=3$  \\
    \hline
         $x^5+2x^3 +x$      &  $q=3$  \\
    \hline
          $x^5 +x^3$      &  $q=3$  \\
    \hline
          $x^5 \pm 2x^3+4x$      &  $q=5$  \\
    \hline

\end{tabular}
\end{center}
\end{table}

\section{Some classes  of CPPs with Niho exponents}

In this section, using Lemmas \ref{lemma} and \ref{lemkey},   we give some classes of CPP exponents of  Niho type.

Let $n=2k $ and $q=p^k.$
Let the conjugate of $x$ over $\F_q$ be denoted by
$\bar{x}$,  i.e., $\bar{x}=x^q.$
Define the unit circle of $\F_{p^n}$ as
 \begin{equation}
U=\{\lambda\in \F_{p^n}|\lambda\bar{\lambda}=1\},\nonumber
\end{equation}
and define the set
\begin{eqnarray}
  \label{aset}
V= \{a|a\in \F_{p^n}, a^{p^k-1}=-1\}.
\end{eqnarray}

\begin{lemma}\cite[Lemma 2]{Nian13}\label{nianlem}
Let $n=2k$ and $d=s(p^k-1)+1,$ where
$s$ is an integer.  Then
the Walsh transform value of
$\Tr_1^n(x^d)$ is given by
$$
(N(a)-1)p^k, \,a\in \F_{p^n},
$$
where
$N(a)$ is the number of $\lambda\in U$ such that
$\lambda^s+\lambda^{1-s}+\bar{a}\lambda +a=0$.
\end{lemma}

\begin{lemma}\label{d3k2lemma}
Let $p=3 $ and $ n=2k$.
Let $N(a)$ be the number of $\lambda\in U$ such that
\begin{eqnarray}\label{p3k2eq}
\lambda^{3^{n-1}}+\lambda^{1-3^{n-1}}+a^{3^k}\lambda+a=0.
\end{eqnarray}
Then $N(a)=1$ if  $a\in V.$
\end{lemma}
{\em Proof:}
Note that for $a\in V,$ we have $a^{3^k}=-a. $ Then from (\ref{p3k2eq}),
$$
\lambda^{3^{n-1}}+\lambda^{1-3^{n-1}}-a\lambda+a=0.
$$
Raising both sides of the above equation to the $3$-th power, we have
$$
\lambda+\lambda^{2}-a^3\lambda^3+a^3=0.
$$
Since $a^3\in V,$ we can replace $a^3$ by $a$,
which leads to
$\lambda+\lambda^{2}-a\lambda^3+a=0. $
This is equivalent to
\begin{eqnarray}
\lambda^3-\lambda^2a^{-1}-\lambda a^{-1}-1=0\label{d3k2eq1}.
\end{eqnarray}
Replacing  $\lambda$ by $\lambda+1$ in (\ref{d3k2eq1}),  we have
$$
\lambda^3a-\lambda^2+1=0,
$$
which becomes
$$
\lambda^3-\lambda+a=0,
$$
by substituting  $\lambda$ with  $\lambda^{-1}$.
 Then to prove that  $N(a)=1$
  is equivalent to show that  the following system of equations have only one root in  $\F_{3^n}$:
\begin{equation}
\left\{\begin{array}{cc}\lambda^3-\lambda+a=0\\
(\lambda^{-1}+1)^{3^k+1}=1,\end{array}\right.\nonumber
\end{equation}
which is equivalent to
\begin{equation}\label{d3k2}
\left\{\begin{array}{cc}\lambda^3-\lambda+a=0\\
\lambda^{3^k}+\lambda+1=0.\end{array}\right.
\end{equation}

It follows from $\lambda^3=\lambda-a$ that
\begin{eqnarray}\label{d3k2eq2}
\lambda^{3^k}=\lambda-\sum_{i=0}^{k-1}a^{3^i}.
\end{eqnarray}
Combining  (\ref{d3k2eq2})  with  the second equation in (\ref{d3k2}), we have
$\lambda=1-\sum_{i=0}^{k-1}a^{3^i}$ is the only solution of (\ref{d3k2}).
Thus $N(a)=1$ for $a\in V.$ This completes the proof.
\done

\begin{theorem}\label{p3kadd2}
Let $p=3 $ and $ n=2k$. Then $ d=3^k+2$ is a CPP
exponent over $\F_{3^n}.$
Moreover, $a^{-1}x^d$ is a CPP over $\F_{3^n}$, where $a\in V,$
and  $V$ is defined by  (\ref{aset}).
\end{theorem}

{\em Proof:}
Note that $n=2k$ and $d=3^k+2=(3^k-1)+3$. This implies $\gcd(d, 3^n-1)=\gcd(3,3^k-1)=1,$ i.e.,
$x^d$ is a PP over $\F_{3^n}$.  In what follows we prove that $x^d+ax$ is also
a PP over $\F_{3^n}$ for $a\in V$.
From Lemma \ref{lemma}, we need to  prove that  for each  $\alpha\in \F_{3^n}^*$,
 $$ \sum_{x\in  \F_{3^{n}}}\omega^{\Tr_1^n(\alpha (x^d+ax))}=0, $$
where $a\in V.$
Since $\gcd(d,3^n-1)=1,$  each nonzero $ \alpha \in \F_{3^n}$ can be written as
$\alpha=t^d$ for a unique $t\in \F_{3^n}^*,$
we have
\begin{eqnarray}
 \sum_{x\in  \F_{3^{n}}}\omega^{\Tr_1^n(\alpha (x^d+ax))}&=&\sum_{x\in  \F_{3^{n}}}\omega^{\Tr_1^n((tx)^d+t^{d-1}atx)}\nonumber\\
 &=& \sum_{x\in  \F_{3^{n}}}\omega^{\Tr_1^n(x^d+t^{d-1}ax)}\nonumber\\
  &=& \sum_{x\in  \F_{3^{n}}}\omega^{\Tr_1^n(x^d+t^{3^k+1}ax)}\label{d3k2eq3}
\end{eqnarray}
 for each $t\in \F_{3^n}^*.$
Since $(t^{3^k+1})^{3^k-1}=1,$ we have $t^{3^k+1}a\in V.$

Let $d'=d\cdot 3^{n-1}=(3^k-1)\cdot 3^{n-1}+3^n\equiv (3^k-1)\times 3^{n-1}+1\,({\rm mod}\, 3^n-1).$
 From Lemma \ref{nianlem},
one has  that for each $a\in \F_{3^n},$
\begin{eqnarray}
\sum_{x\in  \F_{3^{n}}}\omega^{\Tr_1^n(x^d+ax)}
=\sum_{x\in  \F_{3^{n}}}\omega^{\Tr_1^n(x^{d'}+ax)}
=(N(a)-1)\cdot 3^k,\nonumber
\end{eqnarray}
where
$N(a)$ be the number of $\lambda\in U$ such that
$$
\lambda^{3^{n-1}}+\lambda^{1-3^{n-1}}+a^{3^k}\lambda+a=0.
$$
Now let $a\in V,$  from Lemma $\ref{d3k2lemma}$, $N(a)=1.$

Thus
$
\sum_{x\in  \F_{3^{n}}}\omega^{\Tr_1^n(x^d+ax)}=0
$
for all $a\in V.$
Then
from  (\ref{d3k2eq3}), for each $\alpha\in \F^*_{3^n}$,
 $$ \sum_{x\in  \F_{3^{n}}}\omega^{\Tr_1^n(\alpha (x^d+ax))}= \sum_{x\in  \F_{3^{n}}}\omega^{\Tr_1^n(x^d+t^{3^k+1}ax)}=0, $$
due to  $t^{3^k+1}a\in V.$
  This  completes the proof.
\done

\begin{theorem}\label{ppniho2}
Let $p $ be an odd prime and $ n=2k$. Then $ d=(p^k-1)\frac{p^i-1}{2}+p^i$ (for any $1\leq i\leq n$) is a CPP
exponent over $\F_{p^n}.$
Moreover, $a^{-1}x^d$ is a CPP over $\F_{p^n}$, where $a\in V,$
and  $V$ is defined by  (\ref{aset}).
\end{theorem}
{\em Proof:}
Since $\gcd(d,p^k-1)=\gcd(p^i,p^k-1)=1$ and
$\gcd(d,p^k+1)=1$, we have
$\gcd(d,p^{2k}-1)=1.$
So $x^d$ is a PP over $\F_{p^{2k}}$. Next we show that
$x^d+ax$ is a PP
over $\F_{p^{2k}}$ for $a\in V$.

Note that $x^d+ax=x(x^{d-1}+a)=x(x^{(p^k+1)\frac{p^i-1}{2}}+a).$
  Let  $g(x)=x^{\frac{p^i-1}{2}}+a$, then by Lemma \ref{lemkey},
 we only need to  prove  that
$xg(x)^{p^k+1}$ is a PP over $\F_{p^k}$.
Note that
\begin{eqnarray*}
 xg(x)^{p^k+1}&=&x(x^{p^k(\frac{p^i-1}{2})}+a^{p^k})(x^{\frac{p^i-1}{2}}+a)\\
 &=&x(x^{\frac{p^i-1}{2}}-a)(x^{\frac{p^i-1}{2}}+a)\\
 &=&x(x^{p^i-1}-a^2)=x^{p^i}-a^2x.
\end{eqnarray*}
 It is known that a linearized polynomial\footnote{For a prime $p$, a linearized polynomial
over $\F_{p^k}$ is  a polynomial of the form $\sum_{i=0}^{k-1}c_ix^{p^i}$ with coefficients
$c_i$
in
$\F_{p^k}$  \cite{Lidl97}.} $L(x)\in\F_{q}[x]$   is a PP over $\F_{q}$ if
and only if
$x=0$ is its only root in $\F_{q}$ \cite[Theorem 7.9]{Lidl97}.
So that  $x^{p^i}-a^2x$ is a PP over $\F_{p^k}$
if and only if $a^2\neq x^{p^i-1}$ for any $x\in \F^*_{p^k}. $
Suppose that $a^2= x^{p^i-1}$ for some $x\in \F^*_{p^k}, $
then
$(a^2)^{\frac{p^k-1}{2}}=x^{(p^k-1)\frac{p^i-1}{2}}=1,$ which leads to a contradiction
since $a^{p^k-1}=-1$.
 This completes the proof.
\done

\begin{remark}
It can be seen that Theorem
\ref{p3kadd2} is a special case of  Theorem \ref{ppniho2} with $p=3$
and $i=1$. Theorem \ref{p3kadd2} is also a special case of \cite[Corollary 3.4]{Zieve132}.
 The reason we  keep the proof of Theorem
\ref{p3kadd2} is that it gives  a different method to prove  an  exponent
to be a CPP exponent,
and Lemma \ref{d3k2lemma} may be of independent interest.
\end{remark}
%
%

\section{ CPP exponents of  the form $\frac{p^{rk}-1}{p^k-1}+1$ }

Throughout this section, let $p$ be an odd  prime\footnote{The case $p=2$ has been considered in \cite{Wu13}.},  and $n,\,r,\,  k$ be integers such that
$n=rk$.
 Let $d=\frac{p^{rk}-1}{p^k-1}+1,$
where $\gcd(d,p^k-1)=\gcd(r+1,p^k-1)=1.$
From $\gcd(r+1,p^k-1)=1, $
one has that $r$ should be even.


Note that for  $r=2$,  $d=p^k+2$ has been considered in \cite[Corollary 3.4]{Zieve132}.
In this section, we first consider the case  $r=4, $  and  give    sufficient and necessary conditions for
the exponents of the form $
\frac{p^{4k}-1}{p^k-1}+1$ to be  CPP exponents over $\F_{p^n}$. Then we
consider the case $r=6$ with  $p=3$ and $p=5$. Finally,
we  give two    conjectures about some  values of
$r$.

\subsection{$r=4,\,d=\frac{p^{4k}-1}{p^k-1}+1$}\label{p4ksec}
\begin{theorem}\label{ppn4k}
Let $p\neq5$, $r=4$  and $n=4k$.  Let   $d=\frac{p^{4k}-1}{p^k-1}+1$,
where $\gcd(r+1,p^k-1)=\gcd(5,p^k-1)=1.$
Let $a\in\F^*_{p^n}.$
Then $a^{-1}x^d$ is a CPP over $\F_{p^n}$ if  and only if
$a$ satisfies one of  the following  conditions:
\begin{enumerate}
  \item [1)]   $-\frac{2}{5}\lambda_1^2+\lambda_2=0,\,
      -\frac{2}{5}\lambda_1\lambda_3-\frac{3}{125}\lambda_1^4+\frac{3}{25}\lambda_2\lambda_1^2+\lambda_4=0$, and
      $\lambda_3+\frac{4}{25}\lambda_1^3-\frac{3}{5}\lambda_1\lambda_2=0$;

  \item [2)] $p^k\equiv \pm 2\, ({\rm mod }\, 5)$,  $5^{-1}(-\frac{2}{5}\lambda_1^2+\lambda_2)^2=
      -\frac{2}{5}\lambda_1\lambda_3-\frac{3}{125}\lambda_1^4+\frac{3}{25}\lambda_2\lambda_1^2+\lambda_4$, and
      $\lambda_3+\frac{4}{25}\lambda_1^3-\frac{3}{5}\lambda_1\lambda_2=0$;

  \item [3)] $ p=3,$ $k=2,$ $\lambda_2=\lambda_1^2,$ $\lambda_3=-\lambda_1^3$,  and $(\lambda_4+\lambda_1^4)^2=-1$;
  \item [4)]$ p=3,$  $k=1,$ $\lambda_2=\lambda_1^2+1,\, \lambda_3=-\lambda_1^3, $ and $\lambda_4=-\lambda_1^4$;
  \item [5)] $ p=3,$ $k=1,$  $\lambda_2=\lambda_1^2+2,\,\lambda_3=-\lambda_1^3, $ and $\lambda_4=-\lambda_1^4+1$;
  \item [6)] $ p=7,$ $k=1,$     $\lambda_1^2+\lambda_2=0,\,
      \lambda_1\lambda_3+3\lambda_1^4-\lambda_2\lambda_1^2+\lambda_4=0$, and
      $\lambda_3+\lambda_1^3-2\lambda_1\lambda_2=\pm2$;
 \item [7)] $ p=7,$ $k=1,$     $\lambda_1^2+\lambda_2=v,\,
      \lambda_1\lambda_3+3\lambda_1^4-\lambda_2\lambda_1^2+\lambda_4=3v^2$, and
      $\lambda_3+\lambda_1^3-2\lambda_1\lambda_2=\pm1$, where
      $v\in\{3,5,6\}$;
  \item [8)] $ p=13,$ $k=1,$     $-3\lambda_1^2+\lambda_2=v,\,
      -3\lambda_1\lambda_3-2\lambda_1^4-3\lambda_2\lambda_1^2+\lambda_4=3v^2$, and
      $\lambda_3-4\lambda_1^3+2\lambda_1\lambda_2=0$, where
      $v\in\{\pm2,\pm5,\pm6\}$.

\end{enumerate}
 Where $\lambda_i$ is defined in (\ref{hay}), $1\leq i\leq 4$.
\end{theorem}
{\em Proof:}
Since $\gcd(d,p^k-1)=\gcd(5,p^k-1)=1,$ one has that
$x^d$ is a PP over $\F_{p^n}$. In the following we   prove that
$x^d+ax$ is a PP over $\F_{p^k}$
if and only if $a$ satisfies one of  the   conditions in the theorem.

From Lemma \ref{p3nrk}, it is sufficient to prove that
$h_a(x)$ (defined in (\ref{hay})) is a PP over $\F_{p^k}$
if $a$ satisfies one of the conditions in Theorem \ref{ppn4k}.
Recall that
\begin{eqnarray}
h_a(x)= x\sum_{i=0}^{4}\lambda_i x^{4-i}=x^5+\lambda_1x^4+\lambda_2x^3+\lambda_3x^2+\lambda_4x.\label{haypp}
\end{eqnarray}
Replacing $x$ by $x-\frac{\lambda_1}{5}$ in (\ref{haypp}), we get that
\begin{eqnarray}
h_a(x) &=&x^5+(-\frac{2}{5}\lambda_1^2+\lambda_2)x^3+(\lambda_3+\frac{4}{25}\lambda_1^3-\frac{3}{5}\lambda_1\lambda_2)x^2
+(-\frac{2}{5}\lambda_1\lambda_3-\frac{3}{125}\lambda_1^4+\frac{3}{25}\lambda_2\lambda_1^2+\lambda_4)x\nonumber\\
&&-
\frac{1}{5}\lambda_1\lambda_4+\frac{1}{25}\lambda_3\lambda_1^2
+\frac{4}{3125}\lambda_1^5-\frac{1}{125}\lambda_2\lambda_1^3.
\nonumber
\end{eqnarray}
Note that if $h_a(x)-
\frac{1}{5}\lambda_1\lambda_4+\frac{1}{25}\lambda_3\lambda_1^2
+\frac{4}{3125}\lambda_1^5-\frac{1}{125}\lambda_2\lambda_1^3$ is a PP, so does
$h_a(x)$.
Thus, to complete the proof, it is sufficient to show that
 \begin{eqnarray}\label{haypp2}
 h'_{a}(x)=x^5+(-\frac{2}{5}\lambda_1^2+\lambda_2)x^3+(\lambda_3+\frac{4}{25}\lambda_1^3-\frac{3}{5}\lambda_1\lambda_2)x^2
+(-\frac{2}{5}\lambda_1\lambda_3-\frac{3}{125}\lambda_1^4+\frac{3}{25}\lambda_2\lambda_1^2+\lambda_4)x
\end{eqnarray}
 is a PP over $\F_{p^k}$ if $a$ satisfies the
one of the
 conditions in  Theorem \ref{ppn4k}.

%
%

 Then the conditions in the theorem  come from comparing the coefficients of $h'_a(x)$ in
$(\ref{haypp2})$ with the coefficients of the  normalized  PPs of degree $5$ in  Table \ref{ppdeg5}.
We complete the proof.
\done


Let $p=3$ in Theorem \ref{ppn4k}, we have the following corollary.

\begin{corollary}\label{p3n4k}
Let $r=4$ and $n=4k$.  Let   $d=\frac{3^{4k}-1}{3^k-1}+1$,
where $\gcd(r+1,3^k-1)=\gcd(5,3^k-1)=1.$
Let $a\in\F^*_{3^n}.$
Then $a^{-1}x^d$ is a CPP over $\F_{3^n}$ if and only if
$a$ satisfies  the condition $3), $ or $4),$ or $5)$
 in Theorem \ref{ppn4k} or one of the following two conditions:
\begin{itemize}
  \item [1)] $k\equiv 2 \, ({\rm mod}\, 4),  $ $\lambda_2=\lambda_1^2,$ $\lambda_3=-\lambda_1^3, $ and $\lambda_4=-\lambda_1^4$;
  \item [2)]  $k\equiv 1 \, ({\rm mod}\, 2),  $ $\lambda_3=-\lambda_1^3$ and $ -(\lambda_2-\lambda_1^2)^2=(\lambda_4-\lambda_1\lambda_3).$
\end{itemize}
 Where $\lambda_i$ is defined in (\ref{hay}), $1\leq i\leq 4$.
\end{corollary}

Let $n=4k$, where $\gcd(k,4)=1$. Note that
$f(x)=x^4-x-1$ is  an  irreducible polynomial over $\F_{3}$.
 By Lemma \ref{irre},  $f(x)$  remains
irreducible over $\F_{3^k}$.
Let $\beta$ be a root of $x^4-x-1$, then
the order of $\beta$ divides $3^4-1$, thus
$\beta^{80}=1$.
Every  $x\in \F_{3^n} $ can be represented uniquely   as
$x=x_0+x_1\beta+x_2\beta^2+x_3\beta^3,$ where $ x_i\in \F_{3^k}.$

\begin{corollary}\label{p3n4kco}
Let $n=4k$ and  $\gcd(k,4)=1.$ Let $d=\frac{3^{4k}-1}{3^k-1}+1$.
Then $a^{-1}x^d$ is a CPP over $\F_{3^n}$
for the following $a$'s:
\begin{itemize}
    \item  [1)] $a=u(1-\beta^2-\beta^3)+v(\beta+\beta^3);$
    \item  [2)] $a=u(1-\beta^2)+v(\beta-\beta^2-\beta^3);$
    \item  [3)] $a=u(1+\beta)+v(\beta^2-\beta^3); $
    \item  [4)] $a=u(1+\beta^3)+v(\beta+\beta^2).$
\end{itemize}
Where  $u$ and $v$ cannot be zero  at the same time.
\end{corollary}

{\em Proof:}
Let $a=u_0+u_1\beta+u_2\beta^2+u_3\beta^3.$  Then it can be verified that
$a$ satisfied the condition $2)$ in Theorem \ref{p3n4k} if and only if
the following two conditions are satisfied:
\begin{itemize}
  \item [a)] $u_1^3+u_3u_2^2+u_3^2u_2+u_1^2u_2+u_2^3+u_1u_3^2+2u_0u_2^2+u_3^3+2u_0^3+u_3u_1u_0=0$, and
  \item [b)] $u_0^4+2u_1^4+2u_3^4+2u_2^4+2u_1u_3^3+u_2u_3^3+2u_1u_2^3=0.$
\end{itemize}
Then it is readily to verified that all the $a$'s list in the corollary satisfy the conditions a) and b).
This completes the proof.
\done

\begin{remark}
Let $N_k$ denotes the number of $a$'s
such that $a^{-1}x^d$ are  CPPs over $\F_{3^{4k}}$ given in Corollary \ref{p3n4kco}.
Then $N_k=(4\cdot 3^k+2)\cdot (3^k-1)$.
For $k=1,$  $N_1=28,$ and
the Magma results show that  there are  $38$ $a$'s (the extra $10$ $a$'s come from conditions $4)-5)$
in Theorem \ref{ppn4k})  in $\F_{3^4}$ such that $a^{-1}x^d$ are  CPPs over $\F_{3^4}$.
For $k=3,$  $N_3=2860,$ the Magma results show that  these are all the $a$'s such that  $a^{-1}x^d$ are CPPs  over $\F_{3^{4k}}$.
\end{remark}

\begin{remark}
 In Corollary \ref{p3n4k},  for $k=2$,
  there is no $a\in\F_{3^n}$ that
  satisfies condition 1), and
  there are $64$  $a$'s in $\F_{3^{4k}}$ satisfy the condition
 3) in Theorem \ref{ppn4k}, which  are all the $a$'s such that $a^{-1}x^d$ is a CPP over $\F_{3^{4k}}$ according to   the Magma results.
 For $k\geq6$ and  $k\equiv 2 \, ({\rm mod}\, 4),  $ It remains open to give the explicit $a$'s
that   satisfy condition 1).
 \end{remark}

%

In Theorem \ref{ppn4k}, we set $p \neq5$. For $p=5$, we have the following theorem.
\begin{theorem}\label{p5n4k}
Let $r=4$ and $n=4k$.   Let   $d=\frac{5^{4k}-1}{5^k-1}+1$ and $a\in\F^*_{3^n}.$
Then $a^{-1}x^d$ is a CPP over $\F_{5^n}$ if and only if
$a$ satisfies one of the following conditions:
\begin{itemize}
  \item [1)] $\lambda_1=\lambda_2=\lambda_3=0,$ and $-\lambda_4$ is not a fourth power;

  \item [2)] $\lambda_1=0,\, \lambda_2\neq 0,\, -\lambda_2^2=\lambda_4+3\lambda_3^2\lambda_2^{-1},$ and $ 2\lambda_2$ is not a square;

  \item[3)] $k=1$, $\lambda_1=0,$ $\lambda_2=\pm2, $ and $\lambda_4+3\lambda_3^2\lambda_2^{-1}=4$.

\end{itemize}
\end{theorem}

{\em Proof:}
Since $\gcd(d,p^k-1)=\gcd(5,5^k-1)=1,$ one has that
$x^d$ is a PP over $\F_{p^n}$. In the following we   prove that
$x^d+ax$ is a PP over $\F_{p^k}$
if and only if $a$ satisfies one of  the three conditions in the theorem.

From Lemma \ref{p3nrk}, we need to prove that
$h_a(x)$ (defined in (\ref{hay})) is a PP over $\F_{p^k}$
if and only if $a$ satisfies one of the three conditions.
Recall that
\begin{eqnarray}
h_a(x)= x\sum_{i=0}^{4}\lambda_i x^{4-i}=x^5+\lambda_1x^4+\lambda_2x^3+\lambda_3x^2+\lambda_4x.\label{hayp5n4k}
\end{eqnarray}

It is known from Table \ref{ppdeg5} that,   there are
 only five normalized PPs of degree $5$ over $\F_{5^k}$:  $x^5$,
 $x^5-vx$ ($v$ is not a fourth power),
  $x^5-2vx^3+v^2x$ ($v$ is not a square),
 $x^5\pm 2x^3+4x$ ($k=1$).
 Then we discuss the permutation property of $h_a(x)$ as follows:

\textbf{Case \RNum 1:} If $h_a(x)$ is of the form   $x^5-vx$ ($v$ is not a fourth power),  then
 $\lambda_1=\lambda_2=\lambda_3=0$, and
 $-\lambda_4$ is not a fourth power. This is Condition $1)$ in Theorem $\ref{p5n4k}$.

\textbf{Case \RNum 2:}  If $h_a(x)$ is of the form    $x^5-2vx^3+v^2x$ ($v$ is not a square), then one
has that  $\lambda_1=0,\, \lambda_2\neq0$.  Replacing $x$ by $x+3\lambda_3\lambda_2^{-1}$ in  (\ref{hayp5n4k}) we get
 \begin{eqnarray}\label{hayp5n4k2}
 h_a(x)= x^5+\lambda_2x^3+(3\lambda_3^2\lambda_2^{-1} +\lambda_4)x+c^5+\lambda_2c^3+\lambda_3c^2+\lambda_4c,
 \end{eqnarray}
 where $c=3\lambda_3\lambda_2^{-1}.$
 Then $(\ref{hayp5n4k2})$ is a PP over $\F_{5^k}$ if and only if
$ -\lambda_2^2=\lambda_4+3\lambda_3^2\lambda_2^{-1}$ and $ 2\lambda_2$ is not a square.
This is Condition $2)$ in Theorem $\ref{p5n4k}$.

\textbf{Case \RNum 3:} If $h_a(x)$ is of the form   $x^5\pm 2x^3+4x$ ($k=1$),   then
from  $(\ref{hayp5n4k2}),$ we have
$\lambda_1=0,\,\lambda_2=\pm2$, and $\lambda_4+3\lambda_3^2\lambda_1^{-1}=4.$
This is Condition $3)$ in Theorem $\ref{p5n4k}$.

This completes the proof.
\done

\begin{corollary}\label{p5n4kco}
Let $n=4k$ and  $d=\frac{5^{4k}-1}{5^k-1}+1$.
Then $a^{-1}x^d$ is a CPP over $\F_{5^n}$ if
 $a^{2(5^k-1)}=-1$ or   $a^{(5^k-1)}=-1.$
\end{corollary}

{\em Proof:}
It is readily to verify that if $a^{2(5^k-1)}=-1$, then the condition 1)
in Theorem \ref{p5n4k} is satisfied, and
if $a^{5^k-1}=-1$, then the condition 2)
in Theorem \ref{p5n4k} is satisfied.
\done

\begin{remark}
Let $N_k$ denotes the number of $a$'s
such that $a^{-1}x^d$ are  CPPs over $\F_{5^{4k}}$ given in Corollary \ref{p5n4kco}.
Then $N_k=3\cdot(5^k-1)$.
For $k=1,$  $N_1=12,$ and
the Magma results show that  there are  $60$ $a$'s (all of them
satisfy  one of the three conditions in Theorem \ref{p5n4k})
 in $\F_{5^4}$ such that $a^{-1}x^d$ are  CPPs over $\F_{5^4}$.
For $k=2,$  $N_2=72,$
and
 the Magma results show that  there are  $1224$ $a$'s (all of them
satisfy Condition $1)$ or $2)$ in Theorem \ref{p5n4k})   in $\F_{5^8}$ such that $a^{-1}x^d$ are  CPPs over $\F_{5^8}$.
This shows that Corollary \ref{p5n4kco} only gives  a few examples of $a$'s such that $a^{-1}x^d$ are CPPs over
$\F_{5^n}. $
\end{remark}

\subsection{$r=6,\,d=\frac{p^{6k}-1}{p^k-1}+1$}

In \cite{Wu13}, the authors gave some classes of CPP exponents of the form
$d=\frac{2^{rk}-1}{2^k-1}+1$ over $\F_{2^{rk}}$ by using Lemma \ref{p3nrk} and  Dickson polynomials of degree $r+1$
over $\F_{2^k}$, and also in \cite{Zieve132}, the author mentioned that
Dickson polynomials can be used to construct CPPs via Lemma \ref{lemkeyzieve}.
In this subsection, using Lemma \ref{p3nrk} and  Dickson polynomials of degree $7$
over $\F_{p^k}$,  we give some CPP exponents of the form
$d=\frac{p^{6k}-1}{p^k-1}+1$, where  $p=3$ or  $p=5$.

 For a positive integer $l$ and an element  $\eta\in \F^*_{p^k}$, the Dickson polynomial $D_l(x,\eta)$
 over $\F_{p^k}$  is
defined by
$$
D_l(x,\eta)=\sum_{j=0}^{\lfloor l/2\rfloor}\frac{l}{l-j}\binom{l-j}{j}(-\eta)^jx^{l-2j}.
$$
 It is known that a Dickson polynomial is
 a PP over $\F_{p^k}$ if and only if  $\gcd(l,p^{2k}-1)=1$ \cite{Lidl97}.

Let $n=6k$, where $\gcd(k,6)=1$. We know that
$f(x)=x^6+x+2$ is  an  irreducible polynomial over $\F_{3}$.
 By Lemma \ref{irre},  $f(x)$  remains
irreducible over $\F_{3^k}$.
Let $\beta$ be a root of $x^6+x+2$.
 Then every  $x\in \F_{3^n} $ can be represented uniquely   as
$x=x_0+x_1\beta+x_2\beta^2+x_3\beta^3+x_4\beta^4+x_5\beta^5,$ where $ x_i\in \F_{3^k}.$
We represent $x$ by $[x_0,x_1,x_2,x_3,x_4,x_5].$

Note that the  Dickson polynomial  of degree $7$ over $\F_{3^k}$ is  $D_7(x,\eta)=x^7+2\eta x^5+2\eta^2x^3+2\eta^3x$.
The following corollary can be easily proved by using
 Lemma \ref{p3nrk} to the Dickson polynomial $D_7(x,\eta)$  over $\F_{3^k}$.

\begin{corollary}\label{p3n6k}
Let $n=6k$ and  $\gcd(k,6)=1.$ Let $d=\frac{3^{6k}-1}{3^k-1}+1$.
Then $a^{-1}x^d$ is a CPP over $\F_{3^n}$
for the following $a$'s:
\begin{itemize}
    \item  [1)] $a=[0,0,u,u,u,u],[0,u,0,0,u,-u],[0,u,0,-u,-u,0],  $ or  $[u,0,u,0,u,u];$
    \item  [2)] $a=[u,0,-u,u,0,-u],[u,u,0,-u,u,u],[u,u,u,0,-u,0],$ or  $[u,u,u,0,-u,-u];$
    \item  [3)] $a=[u,u,-u,0,-u,-u],[u,u,-u,u,0,0],[u,-u,u,0,0,-u],$ or  $[u,-u,-u,-u,u,-u]. $
 \end{itemize}
Where  $u\neq0$.
\end{corollary}

%

Let $n=6k$, where $\gcd(k,6)=1$. Since
$f(x)=x^6+x+2$ is  an  irreducible polynomial over $\F_{5}$,
 then $f(x)$  is also
irreducible over $\F_{5^k}$  by Lemma \ref{irre}.
Let $\beta$ be a root of $x^6+x+2$.
 Then every  $x\in \F_{5^n} $ can be represented uniquely   as
$x=x_0+x_1\beta+x_2\beta^2+x_3\beta^3+x_4\beta^4+x_5\beta^5,$ where $ x_i\in \F_{5^k}.$

Note that the Dickson polynomial  of degree $7$ over $\F_{5^k}$ is   $D_7(x,\eta)=x^7+3\eta x^5+4\eta^2x^3+3\eta^3x$.
Similarly,
using Lemma \ref{p3nrk} to the  Dickson polynomial $D_7(x,\eta)$ over $\F_{5^k}$,
 we have the following corollary:

\begin{corollary}\label{p5n6k}
Let $n=6k$ and  $\gcd(k,6)=1.$ Let $d=\frac{5^{6k}-1}{5^k-1}+1$.
Then $a^{-1}x^d$ is a CPP over $\F_{5^n}$
for the following $a$'s:
\begin{itemize}
    \item  [1)] $a=[u,0,-u,3u,0,-3u],  [u,u,0,-2u,2u,0], [u,u,0,-u,u,2u], $ or $ [u,u,u,u,u,2u]$;
    \item  [2)] $a=[u,u,2u,-u,0,-u],[u,u,-2u,2u,-u,u], [u,u,-u,0,3u,3u], $ or $ [u,2u,0,u,3u,u]$;
    \item  [3)] $a=[u,2u,u,0,2u,2u], [u,2u,u,2u,u,u], [u,2u,2u,-2u,-u,u], $ or $[u,2u,-2u,-u,2u,u]$;
    \item  [4)] $a=[u,-2u,2u,u,-2u,2u],[u,2u,-u,0,0,u], $ or $ [ u,-u,-2u,2u,-u,2u] $;
    \item  [5)] $a=[0,0,u,u,-2u,0], [0,0,u,3u,-u,-3u],$ or $[0,u,u,u,u,0] $.
 \end{itemize}
Where  $u\neq0$.
\end{corollary}


At the end of this section, we give two  conjectures about the CPP exponents
of the form $d=\frac{p^{rk}-1}{p^k-1}+1.$

\begin{conjecture}\label{conj1}
Let $r+1$ be a prime such that $r+1\neq p$. Let $\gcd(r,k)=1$, $\gcd(r+1,p^2-1)=1$,  and
$d=\frac{p^{rk}-1}{p^k-1}+1$. Then there exist $a's\in \F^*_{p^{rk}}$
such that $h_a(x)$ are Dickson polynomials of degree $r+1$ over $\F_{p^k}$.

\end{conjecture}

This has been proved for $p=2$ with $r=4,\,6,\,10$ (see \cite{Wu13}),  $p=3$ with $r=4,\,6$ (see Corollary \ref{p3n4kco}, Corollary
 \ref{p3n6k}),
and $p=5$ with $r=6$ (see  Corollary \ref{p5n6k}).
Conjecture \ref{conj1} is also verified by Magma for
$p=3$, $r=10$ with $1\leq k \leq 3$.
If Conjecture \ref{conj1} is true, then we have the following proposition:
\begin{proposition}
Let $p,\,r$ and $k$ be defined as  in Conjecture \ref{conj1}.  If Conjecture \ref{conj1} is true, then $d=\frac{p^{rk}-1}{p^k-1}+1$
is a CPP exponent over  over $\F_{p^{rk}}$.
\end{proposition}

{\em Proof:}
Since $r+1 $ is a prime and $r+1\neq p$, we have
$p^r\equiv 1\,({\rm mod}\, r+1)$. Suppose that Conjecture \ref{conj1} is true.  We first show that
$\gcd(r+1,p^{2k}-1)=1$. Assume  that
$p^{2k}\equiv 1\,({\rm mod}\, r+1) $,  together with $p^r\equiv 1\,({\rm mod}\, r+1)$ and $\gcd(k,r)=1$,
one has $p^2\equiv 1\,({\rm mod}\, r+1)$, which is in  contradiction with $\gcd(r+1,p^2-1)=1$.

Since $\gcd(r+1,p^{2k}-1)=1$ and $h_a(y)$ is a Dickson polynomial
of degree $r+1$,  then $h_a(y) $ is a PP over $\F_{p^k}$, which implies
$x^d+ax$ is a PP over $\F_{p^{rk}}$ by  Lemma \ref{p3nrk}.
Then the conclusion follows from $\gcd(d,p^{rk}-1)=\gcd(r+1,p^k-1)=1$.
\done

For $r+1=p$, we have the following conjecture:
\begin{conjecture}\label{conj2}
Let $p$ be an odd prime.
Let $r+1=p$ and $d=\frac{p^{rk}-1}{p^k-1}+1$, then $a^{-1}x^d$ is
a CPP over $\F_{p^{rk}}$, where $a\in \F^*_{p^{rk}}$ such that $a^{p^k-1}=-1$.
\end{conjecture}
This has been proved for  $p=3$ (see \cite[Corollary 3.4]{Zieve132})  and $p=5$  (see Corollary \ref{p5n4kco}),
and also  has  been confirmed by Magma for $p=7$ with $1\leq  k \leq 7$,  $p=11$ with $1\leq  k \leq 3$, and
 $p=13$ with $1\leq  k \leq 3$. If one try to  use Lemma \ref{p3nrk} to tackle Conjecture \ref{conj2},
 the key point is  to prove that
 $x(x^2-a^2)^{\frac{p-1}{2}}$ is a PP over $\F_{p^k}$ for any $k$.

 \begin{remark}
 It can be proved that Conjecture \ref{conj2} is true for  $k=1$. Moreover, we have the following theorem.
 \end{remark}

 \begin{theorem}
Let $p$ be an odd prime.
Let $r+1=p, \,\gcd(rt+1,p-1)=1, $ and $d=t\cdot \frac{p^{r}-1}{p-1}+1$,  then $a^{-1}x^d$ is
a CPP over $\F_{p^{r}}$, where $a\in \F^*_{p^{r}}$ such that $a^{p-1}=-1$.

 \end{theorem}
 {\em Proof:} (Sketch).
 The key point is to show that
 $x(x^{2t}-a^2)^{\frac{p-1}{2}}$ is a PP over $\F_{p}$. Since
 $a^{2(p-1)}=1,$ we have $a^2\in \F_{p}$. Suppose that  $a^2 = x^{2t}$ for some  $x\in\F^*_p$,  we have
 $a^{p-1}=a^{2\cdot\frac{p-1}{2}}=x^{2t\cdot(\frac{p-1}{2})}=1$, which is a contradiction.
 Then we have $a^2 \neq x^{2t}$ for any $x\in\F_p$, which means that
 $(x^{2t}-a^2)^{\frac{p-1}{2}}=\pm 1$ and $x(x^{2t}-a^2)^{\frac{p-1}{2}}=\pm x$.
 Suppose 
 there exist $x_1,\,x_2\in\F_{p}$ such that
 $x_1(x_1^{2t}-a^2)^{\frac{p-1}{2}}=x_2(x_2^{2t}-a^2)^{\frac{p-1}{2}}. $
Then we have $ x_1=\pm x_2$, if $x_1=x_2$, we are done. If $x_1=-x_2$, then $-x_2(x_2^{2t}-a^2)^{\frac{p-1}{2}}=x_2(x_2^{2t}-a^2)^{\frac{p-1}{2}}, $ 
which means $x_1=x_2=0.$
  This completes the proof.
 \done



%
%
%
%
%
%
%
%
%
%
%

\section{A class of multinomial CPPs}
Inspired by \cite{Blokhuis01,Bao13},
in this section,
we give a class of multinomial CPPs, which is a generalization of the main result in
\cite{Bao13}.

\begin{theorem}\label{multith}
Let
$p$ be a prime.  Let $r,\,k$ be two integers with $n=rk$.
  Let  $g(x)$ be a polynomial over $\F_{p^k}$ such that
$xg(x)+vx$ is a PP over $\F_{p^k}$ for some $v\in\F^*_{p^k}$.
Then
$$
f(x)=x( \frac{a}{v}g(\Tr_{k}^{n}(x))+\Tr_{k}^{n}(x)^{p-1})+(p-1)x^{p}+ax
$$
is a CPP over $\F_{p^n}$ if
$\gcd(p-1,r)=1$, $\gcd(r,p)=1,$  and $a\in \F_{p^k}\setminus \{0,-1\}.$
\end{theorem}

{\em Proof:}
We first show that $f(x)$ is a PP over $\F_{p^n}$ if
 $a\in \F_{p^k}\setminus \{0\}.$
Note that
\begin{eqnarray}\label{multieq}
\Tr_{k}^{n}(f(x))&=&\Tr_{k}^{n}(x( \frac{a}{v}g(\Tr_{k}^{n}(x))+\Tr_{k}^{n}(x)^{p-1})+(p-1)x^{p}+ax)\nonumber \\
&=& \frac{a}{v}\Tr_k^n(x)g(\Tr_{k}^{n}(x))+(\Tr_{k}^{n}(x))^{p}+(p-1)\Tr_k^n(x^p)+a\Tr_k^n(x)\nonumber\\
&=& \frac{a}{v}(\Tr_k^n(x)g(\Tr_{k}^{n}(x))+v\Tr_k^n(x))
\end{eqnarray}
for all $x\in \F_{p^n}.$
Suppose there exists  $x,\, y\in\F_{p^n}$ such that   $f(x)=f(y)$. Then  from $(\ref{multieq})$, one has
$$\frac{a}{v}(\Tr_k^n(x)g(\Tr_{k}^{n}(x))+v\Tr_k^n(x))=\frac{a}{v}(\Tr_k^n(y)g(\Tr_{k}^{n}(y))+v\Tr_k^n(y)). $$
Thus  $\Tr_k^n(x)=\Tr_k^n(y)$ due to $xg(x)+vx $ is a PP over $\F_{p^k}.$
Let   $\Tr_k^n(x)=\Tr_k^n(y)=\alpha,$ from $f(x)=f(y)$,   we have
$$(\frac{a}{v}g(\alpha)+\alpha^{p-1}+a)x+(p-1)x^p=(\frac{a}{v}g(\alpha)+\alpha^{p-1}+a)y+(p-1)y^p,$$
i.e.,   $(\frac{a}{v}g(\alpha)+\alpha^{p-1}+a)(x-y)=(x-y)^p.$
If $x\ne y,$ then we have
$(x-y)^{p-1}=\frac{a}{v}g(\alpha)+\alpha^{p-1}+a\in\F_{p^k}.$ Let
the order of $x-y$ be $h$, we have
$$h|(p-1)(p^k-1) \,{\rm and } \,h|p^n-1,$$
which implies $h|\gcd((p-1)(p^k-1),p^{rk}-1)$.
Since $\gcd(p-1,\frac{p^{rk-1}}{p^{k}-1})=\gcd(p-1,r)=1$,
we have $\gcd((p-1)(p^k-1),p^{rk}-1)=p^k-1$ and $h|p^k-1,$
which means that $x-y\in \F_{p^k}.$
Since $\Tr_k^n(x)=\Tr_k^n(y)$, we have $\Tr_{k}^n(x-y)=r(x-y)=0$, then it  follows from $\gcd(r,p)=1$ that
$x=y$.
This shows that  $f(x)$ is a PP over $\F_{p^{n}}$ if $a\in \F_{p^k}\setminus \{0\},$
which implies that
$f(x)$ is a CPP over $\F_{p^{n}}$ if $a\in \F_{p^k}\setminus \{0,-1\}.$
This completes  the proof.
\done

In the following, we give some explicit $g(x)$ such that $xg(x)+vx\in\F_{p^k}[x]$ are  PPs  over
$\F_{p^k}$, then according to Theorem \ref{multith}, we can obtain  some classes  of CPPs over $\F_{p^{rk}}$.

The simplest choice is  $g(x)=0$. Fixing  $g(x)=0$,  we have the following corollary.

\begin{corollary}\label{multicoro}
 Let
$p$ be a prime, and  let $r,\,k$ be two integers with $n=rk$.
  Let $a\in \F_{p^k}\setminus \{0,-1\}.$ Then
$$
f(x)=x(\Tr_{k}^{n}(x))^{p-1}+(p-1)x^{p}+ax
$$
is a CPP over $\F_{p^n}$ if
$\gcd(p-1,r)=1$ and $\gcd(r,p)=1.$
\end{corollary}

\begin{remark}
In Corollary \ref{multicoro}, let  $p=2$ and $r$ be odd, then $\gcd(p,r)=\gcd(p-1,r)=1,$ we have
$f(x)=x\Tr_k^n(x)+x^2+ax$,
which are  the  CPPs given in \cite[Corollary 2.2]{Bao13}.
\end{remark}

Let $g(x)=x^{d-1}\in \F_{p^k}[x],$ where $d$ is a CPP exponent over $\F_{p^k}$.
Then
there exists $v\in \F^*_{p^k}$ such that
$xg(x)+vx$ is a PP over $\F_{p^k}.$ By Theorem \ref{multith}, we  can obtain many  classes of CPPs over $\F_{p^{rk}}.$
All the CPP exponents proposed in this paper  and in \cite{Tu13,Tu132,Wu13,Zieve13} can be used in Theorem \ref{multith}.

\begin{example}
From Theorem \ref{p3kadd2}, $d=3^k+2$ is a CPP exponent over $\F_{3^{2k}}$. Let
$g(x)=x^{d-1}$ and $v\in V$, where $v$ is defined in (\ref{aset}). Then $xg(x)+vx$ is a PP
over $\F_{3^{2k}}$.
The function  defined in
Theorem \ref{multith} is given by
$$
f(x)=x[\frac{a}{v}((\Tr_k^{2kr}(x))^{3^k+1})+(\Tr_k^{2kr}(x))^2]+2x^3+ax,
$$
and it is a CPP over  $\F_{3^{2kr}}$ if   $\gcd(r,6)=1$ and $a\in \F_{3^{2k}}\setminus \{0,-1\}$.
\end{example}

One can also  choose $g(x)$ such that $xg(x)+vx$ is a Dickson polynomial of degree $l$ over $\F_{p^k}$.
\begin{example}
Assume $g(x)=x^4+c_0x^2+c_1\in\F_{3^k}[x]$. If $2c_0^2=c_1+v, $ then  $xg(x)+vx=x^5+c_0x^3+(c_1+v)x$ is  a Dickson polynomial
of degree $5$.
Suppose  $\gcd(5,3^{2k}-1)=1,$
 we have  $xg(x)+vx$ is a PP over $\F_{3^k}$. Then
the function  defined in
Theorem \ref{multith} is given by
$$
f(x)=x[\frac{a}{v}((\Tr_k^{rk}(x))^4+c_0(\Tr_k^{rk}(x))^2+c_1)+(\Tr_k^{rk}(x))^2]+2x^3+ax,
$$
and it is a CPP over  $\F_{3^{rk}}$ if   $\gcd(r,6)=1$ and  $a\in \F_{3^{k}}\setminus \{0,-1\}$.
\end{example}

\end{document}